\documentclass[aps,prl,twocolumn,showpacs,superscriptaddress,groupedaddress]{revtex4-1}  % for review and submission
\usepackage{graphicx}  % needed for figures
\usepackage{dcolumn}   % needed for some tables
\usepackage{bm}        % for math
\usepackage{amssymb}   % for math
\usepackage{amsmath}   % for math

\def\pd#1#2{\frac{\partial #1}{\partial #2}}

\def\d{\mathrm{d}}

\RequirePackage[normalem]{ulem}
\RequirePackage{color}\definecolor{myred}{rgb}{0.6,0,0}\definecolor{myblue}{rgb}{0,0,0.6}

\providecommand{\deleted}[1]{{}}

\begin{document}

\title{Recent progress on phase-space turbulence and dynamical response in collisionless plasmas}
\author{M.~Lesur$^1$}
\affiliation{$^1$Itoh Research Center for Plasma Turbulence, Kasuga, Kasuga Koen 6-1, 816-8580, Kyushu University, Japan}

\date{\today}

\begin{abstract}
In the presence of wave dissipation, phase-space structures emerge in nonlinear Vlasov dynamics. Their dynamics can lead to a nonlinear continuous shifting of the wave frequency (chirping).
This report summarizes my personal contribution to these topics in the fiscal year 2012.
The effects of collisions on chirping characteristics were investigated, with a one-dimensional beam-plasma kinetic model.
The long-time nonlinear evolution was systematically categorized as damped, steady-state, periodic, chaotic and chirping. The chirping regime was sub-categorized as periodic, chaotic, bursty, and intermittent.
Existing analytic theory was extended to account for Krook-like collisions. Relaxation oscillations, associated with chirping bursts, were investigated in the presence of dynamical friction and velocity-diffusion. The period increases with decreasing drag, and weakly increases with decreasing diffusion.
A new theory gives a simple relation between the growth of phase-space structures and that of the wave energy. When dissipation is modeled by a linear term in the field equation, simple expressions of a single hole growth rate and of the initial perturbation threshold are in agreement with numerical simulations.
\end{abstract}

% 05.65.+b	Self-organized systems
% 07.05.Tp	Computer modeling and simulation
% 41.20.Cv	Electrostatics; Poisson and Laplace equations, boundary-value problems
% 47.27.Cn	Transition to turbulence
% 47.27.De	Coherent structures
% 47.27.E-	Turbulence simulation and modeling
% 47.27.eb	Statistical theories and models
% 47.27.ed	Dynamical systems approaches
% 47.27.ef	Field-theoretic formulations and renormalization
% 47.27.ek	Direct numerical simulations
% 47.27.T-	Turbulent transport processes
% 47.27.tb	Turbulent diffusion
% 52.25.Dg	Plasma kinetic equations
% 52.25.Fi	Transport properties
% 52.25.Gj	Fluctuation and chaos phenomena (for plasma turbulence, see 52.35.Ra; see also 05.45.-a Nonlinear dynamics and chaos)
% 52.25.Xz	Magnetized plasmas
% 52.30.Gz	Gyrokinetics
% 52.35.Bj	Magnetohydrodynamic waves (e.g., Alfven waves)
% 52.35.Fp	Electrostatic waves and oscillations (e.g., ion-acoustic waves)
% 52.35.Kt	Drift waves
% 52.35.Mw	Nonlinear phenomena: waves, wave propagation, and other interactions (including parametric effects, mode coupling, ponderomotive effects, etc.)
% 52.35.Py	Macroinstabilities (hydromagnetic, e.g., kink, fire-hose, mirror, ballooning, tearing, trapped-particle, flute, Rayleigh-Taylor, etc.)
% 52.35.Qz	Microinstabilities (ion-acoustic, two-stream, loss-cone, beam-plasma, drift, ion- or electron-cyclotron, etc.)
% 52.35.Ra	Plasma turbulence
% 52.35.Sb	Solitons; BGK modes
% 52.55.Fa	Tokamaks, spherical tokamaks
% 52.65.Ff	Fokker-Planck and Vlasov equation
% 52.65.Tt	Gyrofluid and gyrokinetic simulations
\maketitle

% To show line number every 5 lines
%\setpagewiselinenumbers
%\modulolinenumbers[5]
%\linenumbers

The nonlinear evolution of collisionless or weakly collisional plasmas is often accompanied by the formation and ballistic propagation of self-trapped vortices in phase-space \cite{schamel86}. These coherent structures are spontaneously formed in phase-space by resonant nonlinear wave-particle interactions, which trap particles in a trough. These trapped particles in turn generate a self-potential, leading to a self-sustained structure, which can break ties from resonance. Self-trapped structures resemble vortices in 2D ideal fluid turbulence \cite{mcwilliams84}, and so reflect similarities between 2D fluids governed by Euler's equation and 1D plasmas governed by Vlasov's equation.
Both these systems conserve phase-space density. The evolution of each system is constrained by two invariants: energy and enstrophy in the fluid case, and energy and total phasestrophy $\int f^2 \d x \d v$ in the Vlasov case \cite{lesur13pre}. Phase-space vortices, called as electron and ion holes \cite{schamel12}, open a new channel for tapping free energy in the system. They are key nonlinear agents of instability \cite{dupree82}, anomalous transport \cite{hsu72} and anomalous resistivity \cite{dupree70}, in the context of energetic particle-driven activities in space and magnetic fusion plasmas \cite{eliasson06}, collisionless magnetic reconnection \cite{drake03}, collisionless shock waves \cite{sakanaka72} and alpha-channeling \cite{mynick94}.
Much progress in the still-evolving topic of kinetic nonlinearities is based on the paradigm of phase-space turbulence \cite{diamonditoh2,diamond_tutorial,kosuga11}, where the system is treated as an ensemble of structures in phase-space, rather than an ensemble of waves, as in quasi-linear theory.
We can contrast phase-space turbulence with conventional approaches in terms of the Kubo number $K=\omega _b \tau _c$, which measures the coherence of turbulence. Here, $\omega _b$ is the bounce frequency of trapped particles, and $\tau _c$ is the correlation time of a structure. Conventional theories that rely on linear waves and their nonlinear extensions (mode coupling, weak and strong turbulence theories) require $K<1$ for their validity. This condition is easily violated when wave-particle interactions are strong. Phase-space turbulence theory concerns the ubiquitous $K\gtrsim 1$ regime.
Existing phase-space turbulence literature focuses on the instability and dynamics of phase-space holes \cite{roberts67,berk70,saeki79,schamel79,dupree82,berman85,berkbreizman97pla,lesur09,lilley10,nyqvist12} and granulations \cite{dupree72,boutrosghali82}.

\section{Models}

The simplest model for describing phase-space hole dynamics is the Berk-Breizman (BB) extension of the bump-on-tail instability. It includes a finite, fixed wave damping ($\gamma _d$), and a collision operator with drag ($\nu _f$) and diffusion ($\nu _d$). We adopt a perturbative approach, and cast the BB model in a reduced form, which describes the time evolution of the beam particles only \cite{berkbreizman99, lesur12}. In this sense, we note that the BB model with extrinsic dissipation is also applicable to the traveling wave tube "quasilinear experiment" with a lossy helix \cite{tsunoda87}.
In this model, a single electrostatic wave with a wave number $k$ is assumed and the real frequency of the wave is set to $\omega = \omega _p$, the Langmuir plasma frequency.
The evolution of the beam distribution, $f(x,v,t)$, is given by a kinetic equation,
%\begin{equation}
%\begin{split}
%\pd{f}{t} \,+\, v \, \pd{f}{x} \,+\, \frac{q E}{m} &\, \pd{f}{v} \;=\; \\
%& -\nu _a \delta f \,+\, \frac{\nu _f^2}{k} \frac{\partial \delta f}{\partial v} \,+\, \frac{\nu _d^3}{k^2} \frac{\partial ^2 \delta f}{\partial v^2}, \label{model:bb_kinetic}
%\end{split}
%\end{equation}
\begin{equation}
\pd{f}{t} + v  \pd{f}{x} + \frac{q E}{m}  \pd{f}{v} \,=\, -\nu _a \delta f + \frac{\nu _f^2}{k} \frac{\partial \delta f}{\partial v} + \frac{\nu _d^3}{k^2} \frac{\partial ^2 \delta f}{\partial v^2}, \label{model:bb_kinetic}
\end{equation}
where $\delta f\equiv f-f_0$, $f_0(v)$ is the initial velocity distribution, and $\nu _a$, $\nu _f$, $\nu _d$ are constants characterizing Krook-type collisions, velocity-space diffusion and dynamical friction, respectively.
The evolution of the pseudo-electric field $E\equiv Z \exp{i\zeta} + c.c.$ is given by
\begin{equation}
\frac{\mathrm{d} Z}{\mathrm{d} t} \;=\; - \frac{m \omega _p^3}{4\pi q n_0} \, \int f(x,v,t) \, e^{-i \zeta} \, \mathrm{d}x \, \mathrm{d}v \;-\; \gamma _d \, Z, \label{model:reduced_Z}
\end{equation}
where $\zeta \equiv k x-\omega t$, and $n_0$ is the total density.

For the ion-acoustic wave model, we include two species $s=i,e$, assume collisions are negligible, and do not filter a particular wave number. The CDIA model is composed of two kinetic equations,
\begin{equation}
\pd{f_s}{t} \,+\, v \, \pd{f_s}{x} \,+\, \frac{q_s E}{m_s} \, \pd{f_s}{v} \;=\; 0, \label{model:cdia_kinetic}
\end{equation}
and a current equation,
\begin{equation}
\frac{\partial E}{\partial t} \;=\; \frac{m \omega _p^2}{n_0q} \, \sum _s \int v f_s(x,v,t) \, \mathrm{d}v. \label{model:cdia_current}
\end{equation}

\section{Nonlinear categorization of the beam-driven instability}

In Ref.~\cite{lesur12}, we applied the BB model to one dimensional plasma, to investigate the kinetic nonlinearities, which arise from the resonance of a single electrostatic wave with an energetic particle beam. We developed a systematic categorization of the long-time nonlinear evolution as damped, steady-state, periodic, chaotic and chirping.
A similar categorization had been performed numerically \cite{vann03, lesur09}, in the case where collisions are modelled by a simple Krook-like operator with a collision frequency $\nu _a$.
However, Lilley and Lesur have shown that the inclusions in the collision operator of dynamical friction, or drag, and diffusion have a strong impact on the nonlinear behaviour, and is necessary to qualitatively reproduce experimental chirping AEs \cite{lilley09, lesur10}.

Using our Vlasov code COBBLES Ref.~\cite{lesur09,lesur12,lesur07}, we scanned the parameter space for a fixed value of the linear drive normalized to the linear frequency, $\gamma _{L0}/\omega_0=0.1$. We developed sub-categories for the chirping regime, as periodic, chaotic, bursty, and intermittent. Up-down asymmetry and hooked chirping branches are also categorized. For large drag, we observed holes with quasi-constant velocity, in which case the solution is categorized into steady hole, wavering hole and oscillating hole. We considered two complementary parameter spaces: 1.~the ($\gamma _d$, $\nu _d$) space for fixed $\nu _d/\nu _f$ ratios; 2.~the ($\nu _f$, $\nu _d$) space for fixed $\gamma _d/\gamma _{L0}$ ratios, close to and far from marginal stability. The presence of drag and diffusion (instead of a Krook model) qualitatively modifies the nonlinear bifurcations. The bifurcations between steady-state, periodic and steady hole solutions agree with analytic theory. Moreover, the boundary between steady and periodic solutions agree with analytic theory. We observed nonlinear instabilities in both subcritical and barely unstable regime. We showed that quasi-periodic chirping is a special case of bursty chirping, limited to a region relatively far from marginal stability.

\begin{figure} \begin{center}
\includegraphics{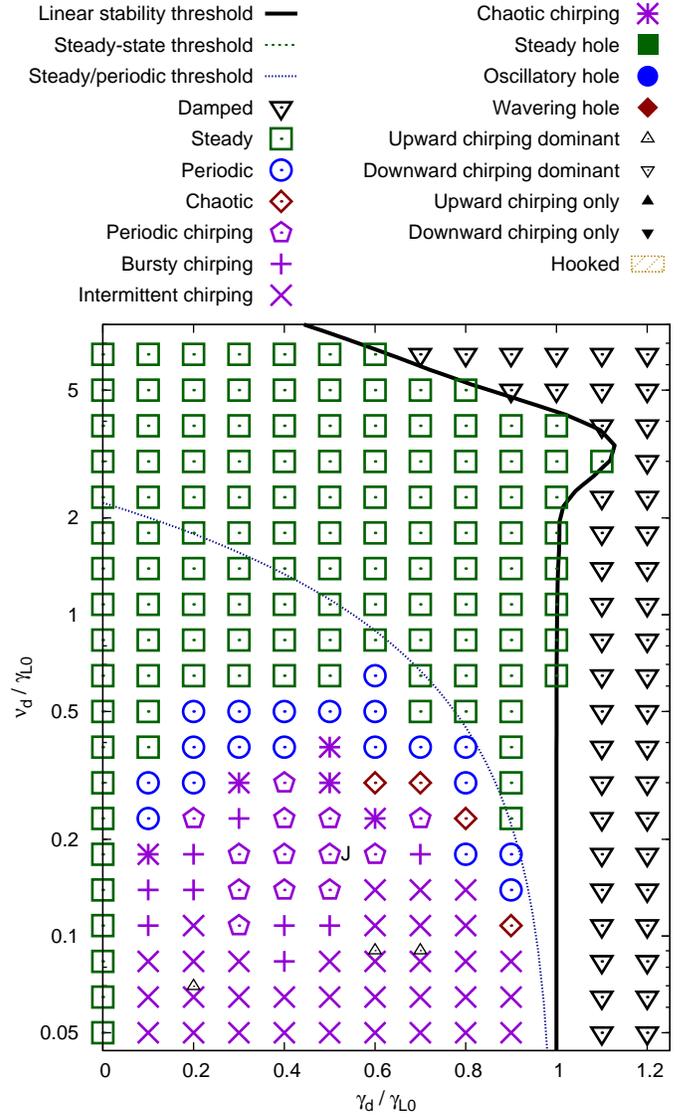}
\caption{Behaviour bifurcation diagram for $\gamma _{L0}=0.1$ and $\nu_d / \nu_f = 5$. The classification of each solution is plotted in the ($\gamma _d$,\, $\nu _d$) parameter space. The legend is shared between Figs.~\ref{fig:bifurcation_R5} and \ref{fig:bifurcation_R1}. The letter J indicates the JT-60U discharge E32359.}
\label{fig:bifurcation_R5}
\end{center} \end{figure}

Fig.~\ref{fig:bifurcation_R5} shows the categorization of each simulation result in the ($\gamma _d$,\, $\nu _d$) parameter space, in the small-drag regime, $\nu_f \ll \nu_d$. The phase diagram is qualitatively similar to what was obtained with Krook collisions, although chirping solutions can be intermittent, bursty or periodic, in addition to the chaotic behaviour found in the Krook case. We showed that quasi-periodic chirping is a special case of bursty chirping, limited to a region where $\gamma _d / \gamma _{L0}=0.2-0.7$.

\begin{figure} \begin{center}
\includegraphics{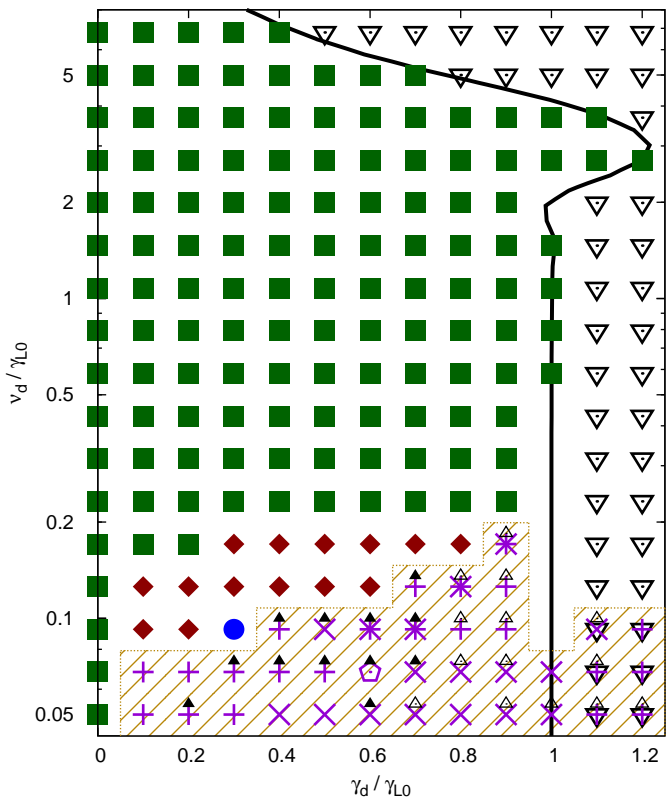}
\caption{Behaviour bifurcation diagram for $\gamma _{L0}=0.1$ and $\nu_d / \nu_f = 1$. The classification of each solution is plotted in the ($\gamma _d$,\, $\nu _d$) parameter space. In the bottom right corner, superposed symbols show subcritical solutions for which the amplitude is damped when $\omega _b/\gamma _{L0}=0.03$, but unstable when $\omega _b/\gamma _{L0}=1$.}
\label{fig:bifurcation_R1}
\end{center} \end{figure}

Fig.~\ref{fig:bifurcation_R1} shows the categorization of each simulation result in the ($\gamma _d$,\, $\nu _d$) parameter space, in the large-drag regime, $\nu_d / \nu_f = 1$.
The presence of significant drag qualitatively modifies the nonlinear bifurcations. Steady-state, periodic and chaotic solutions, which are devoid of significant phase-space structure dynamics, are replaced by long-lived phase-space holes. The periodic chirping regime almost disappears.
We confirmed that steady-state solutions only exist for collision frequency above a threshold predicted by analytic theory \cite{lilley09}. Moreover, the boundary between steady and periodic solutions agree with analytic theory when the system is close to marginal stability.
We found nonlinear instabilities in both subcritical ($\gamma<0$) and barely unstable ($\gamma \ll \gamma _{L0}$) regime.

We don't observe neither long-lived clump or down-chirping dominant cases, which are often observed in the experiment. Although the reason eludes us, we can speculate that down-chirping dominant cases in TAE experiments may be the result of reversed magnetic shear, which effectively brings a minus sign in front of the drag term in Vlasov equation when the 3D Fokker-Planck operator is projected on the resonant surface. However, there are other possible causes, such as particular shapes of $f_0$ with non-constant slope, departures from the adiabatic bulk hypothesis, and processes that are not included in the BB model.

\section{Effects of collisions on energetic particle-driven chirping bursts}

A feature of the nonlinear evolution of AEs, the frequency sweeping (chirping) of the resonant frequency by 10-30$\%$ on a timescale much faster than the equilibrium evolution, has been observed in the plasma core region of tokamaks JT-60U \cite{kusama99}, DIII-D \cite{heidbrink95}, the Small Tight Aspect Ratio Tokamak (START) \cite{mcclements99}, the mega amp spherical tokamak (MAST) \cite{pinches04S47}, the National Spherical Torus Experiment (NSTX) \cite{fredrickson06}, and in stellerators such as the Compact Helical Stellerator (CHS) \cite{takechi99}. In general, two branches coexist, with their frequency sweeping downwardly (down-chirping) for one, upwardly (up-chirping) for the other. In most of the experiments, relaxation oscillations are observed, with quasi-periodical chirping bursts. The period is in the order of the millisecond. Chirping bursts are associated with significant transport of energetic particles, in particular when they trigger avalanches \cite{medley04}. This motivates our investigation of the period of chirping bursts.

\subsection{Effect of Krook collision on chirping velocity}

\begin{figure} \begin{center}
\includegraphics{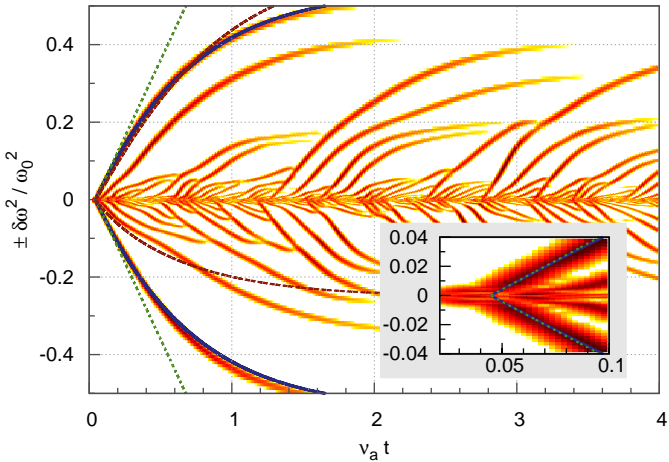}
\caption{Effect of finite Krook collisions on chirping velocity. Spectrogram of the electric field. Logarithmic color code ranging from $1$ (black) to $10^{-3}$ (white). Two dotted, straight lines correspond to $\delta \omega \sim \sqrt{t}$. Two solid curves correspond to Eq.~(\ref{eq:chirping_velocity_with_krook}). We include a correction coefficient $\beta=1.23$. Two dashed curves correspond to Ref.~\cite{nyqvist12}. Inset: zoom on the beginning of the first chirping event.}
\label{fig:finite_nua_spectrum}
\end{center} \end{figure}

In Ref.~\cite{lesur13pop}, I investigated the effects of collisions on chirping characteristics, with a one-dimensional kinetic model.
Existing theory predicts the time evolution of the frequency shift as $\delta \omega \sim \sqrt{t}$ \cite{berkbreizman97pla}. In this work, we extend the latter theory by accounting for Krook-like collisions with frequency $\nu_a$, which yields
\begin{eqnarray}
\delta \omega (t) &=& \pm \alpha \, \beta \, \gamma _{L0} \,\sqrt{\gamma _d t} \, \left[ 1\,-\, \frac{1}{3} (\nu _a t) \,+\, \frac{7}{90} (\nu _a t)^2 \right. \nonumber \\
& & \left. \,-\, \frac{19}{1890} (\nu _a t)^3 \,+\, \frac{1507}{1701000} (\nu _a t)^4 \,+\, \ldots \right]. \label{eq:chirping_velocity_with_krook}
\end{eqnarray}
This is consistent with numerical simulations, where chirping significantly departs from the widely-accepted square-root time dependency.

\section{Effect of drag and diffusion on chirping period}

I investigated relaxation oscillations, which are associated with experimental observations of chirping bursts. These oscillations are recovered in simulations in the presence of dynamical friction and velocity-diffusion.
The period increases with decreasing drag, and weakly increases with decreasing diffusion.

\begin{figure} \begin{center}
\includegraphics{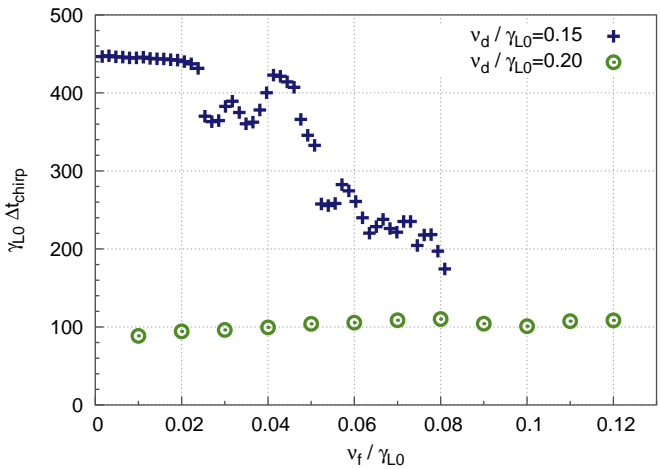}
\caption{Effect of drag on chirping period at fixed diffusion, for $\nu _d = 0.015$ and $0.020$.}
\label{fig:Period_nuf}
\end{center} \end{figure}

The effect of drag on chirping period is complex and depends on other parameters. Fig.~\ref{fig:Period_nuf} shows the period as a function of drag, for two fixed values of diffusion. The data points are shown only for simulations categorized as periodic chirping, by the categorization algorithm developed in Ref.~\cite{lesur12}. We observe that, when the period is large, the general trend is a decreasing period as drag increasing. The trend is reversed when the period is small.

\begin{figure} \begin{center}
\includegraphics{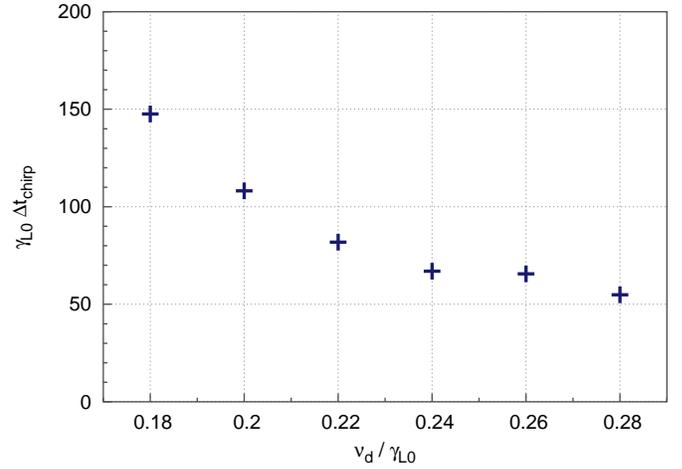}
\caption{Effect of diffusion on chirping period at fixed drag, $\nu _f = 0.005$.}
\label{fig:Period_nud}
\end{center} \end{figure}

The effect of diffusion on chirping period is consistent with simple intuitive arguments. Fig.~\ref{fig:Period_nud} shows that the period decreases as diffusion increases, for fixed drag. For large diffusion, the effect tends to saturate. Again, the data points are shown only for simulations categorized as periodic chirping.

\begin{figure} \begin{center}
\includegraphics{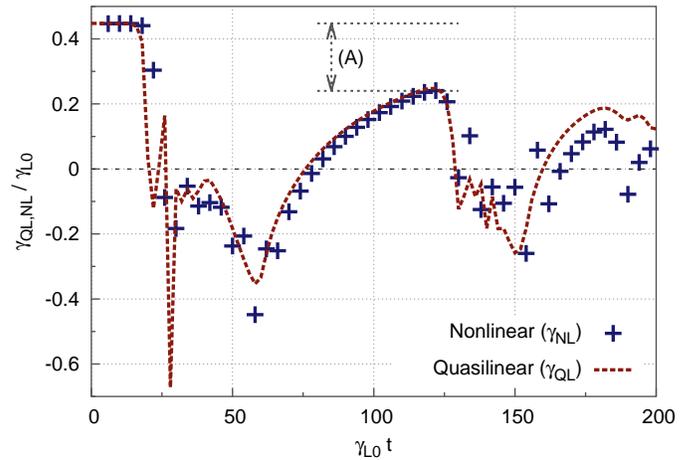}
\caption{Time-evolution of the growth rate in our reference case. Points: obtained from the time-series of electric field amplitude. Dashed curve: obtained from the linearized equations and the Gaussian model for $\partial \left\langle f \right\rangle / \partial v$. In the absence of hole and clump, $\gamma / \gamma_{L0} = 0.4476$. (A) shows the discrepancy between the latter value and the maximum growth rate reached before the second burst.}
\label{fig:Growth_rate_recovered}
\end{center} \end{figure}

\begin{figure} \begin{center}
\includegraphics{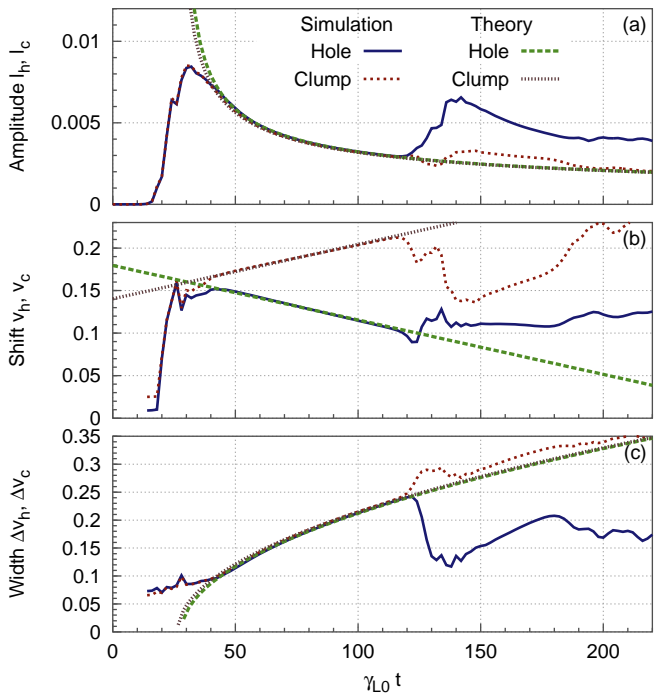}
\caption{Time-evolution of hole and clump characteristics, for our reference case ($\nu_f=0.008$, $\nu_d=0.025$). (a) Amplitude $l_h$ and $l_c$. (b) Shift $v_h$ and $v_c$. (c) Width $\Delta v_h$ and $\Delta v_c$. The dashed curves correspond to theory, with $\gamma _{L0} t_0 = 100$.}
\label{fig:HoleClumpEvolution}
\end{center} \end{figure}

I clarified the mechanism with a simple semi-analytic model of hole/clump pair. Between two bursts, the wave amplitude is low, and collisions dominate over the nonlinear term in the kinetic equation. By modeling a hole and a clump in the velocity distribution by two Gaussians, their dynamics is obtained as the analytic solution of a Fokker-Planck equation, given an initial fit of the structures just after a burst. Fig.~\ref{fig:Growth_rate_recovered} shows that the quasi-linear growth rate obtained from the eigenvalue problem with the Gaussian model for hole and clump is in good agreement with the nonlinear growth rate extracted from the simulation. There is also a good agreement for the details of hole/clump width, amplitude and shift, as shown in Fig.~\ref{fig:HoleClumpEvolution}. The instantaneous quasi-linear growth rate was then obtained numerically by solving a linear equation system. This procedure recovers time-evolution of amplitude growth and leads to a better qualitative understanding of the nonlinear evolution of wave amplitude between bursts.

Let me make an important remark. Fig.~\ref{fig:HoleClumpEvolution}(a) shows that the second burst occurs before the remnant hole-clump pair from the first burst is completely dissipated. Since the initial distribution function is not recovered, there is a discrepancy between the linear growth rate $\gamma=0.04476$ and the maximum growth reached before the second burst at $\gamma _{L0} t = 122$, $\gamma_{NL}=0.024$. This discrepancy is marked (A) in Fig.~\ref{fig:Growth_rate_recovered}.
In Alfv{\'e}n waves experiments in magnetic confinement devices, the amplitude time-series of magnetic perturbation looks as though the linear growth rate $\gamma$ can be extracted by fitting an exponential to the signal. Our analysis shows that this procedure can lead to large error ($50\%$ in our case). In other words the growth is not linear in the case of quasi-periodic chirping bursts.
For the same reason, successive chirping rates may not reflect the relaxed distribution ($f_0(v)$ in the collision operator $\mathcal{C}(f-f0)$), but the instantaneous state of the relaxing distribution ($f(x,v,t)$).
In addition, since the discrepancy $(A)$ depends on the details of the velocity distribution, theory predicts the timing of the subsequent burst (e.g. at $\gamma _{L0} t \approx 125$), but only qualitatively.

\section{Nonlinear instabilities driven by coherent phase-space structures}

Instability dynamics \cite{chandrasekhar,manneville} is of great interest in the context of pattern formation \cite{cross93}, the onset of turbulence \cite{landau44}, and many other subjects. While instabilities are central to virtually every field of physics, in collisionless or weakly collisional plasmas the disparate roles of resonant and non-resonant particles offers an interesting variation on time-honored methods and approaches. In this respect, it has long been realized that wave and instability dynamics and evolution in a collisionless plasma can be described in terms of coupled, inter-penetrating ensembles of resonant and non-resonant particles or equivalently, resonant particles and a gas of plasmon quasi-particles. While the linear theory of the Vlasov plasma is well established, its nonlinear theory is a rich and still-evolving subject. Rather little, however, is understood about nonlinear, or subcritical, Vlasov stability, in which the growth process circumvents linear theory \cite{schamel12}. One idea concerning subcritical processes derives from the properties of phase-space granulations or structures, which can exchange momentum via channels which differ from that of familiar wave-particle resonance, and so can tap free energy when wave excitation cannot \cite{dupree82}. Such granulations are self-bound aggregations of resonant particles, which constitute a novel collective exciton.

\begin{figure}
\includegraphics{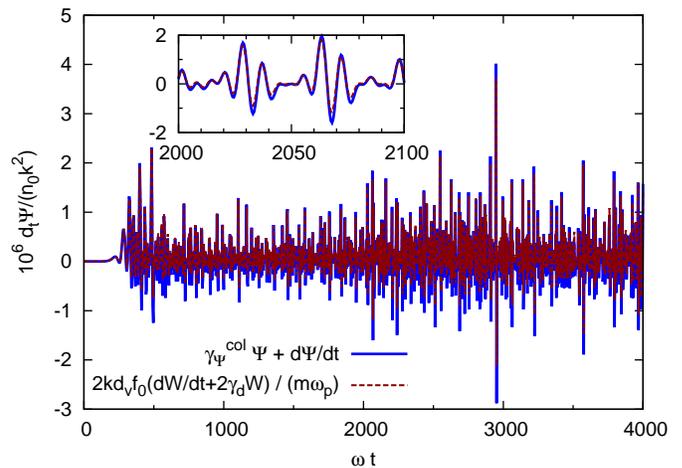}
\caption{\label{fig:bb_ps_growth}Growth of phasestrophy and wave energy in the BB case. Inset: zoom on a smaller timescale. Simulation parameters are $\gamma _{L0} / \omega =0.1$, $\gamma _d / \gamma _{L0}=0.7$, $\nu _a / \gamma _{L0} = 10^{-3}$ and $\nu _f=\nu _d = 0$.}
\end{figure}

\begin{figure}
\includegraphics{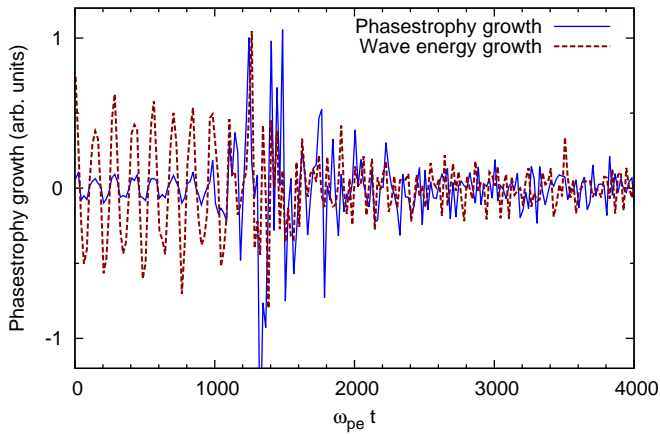}
\caption{\label{fig:cdia_ps_growth}Growth of phasestrophy and wave energy in the CDIA case.}
\end{figure}

In Refs.~\cite{lesur13pre,lesur12fec,lesur12aps,lesur13iaea}, we presented a new theory of subcritical Vlasov plasma instability formulated in terms of the evolution of waves and phase-space density correlations.
The evolution of phase-space structures follows that of the phasestrophy \cite{diamonditoh2,kosuga11},
\begin{equation}
\Psi _s \;\equiv\; \int _{-\infty} ^{\infty} \left\langle \delta f _s ^2 \right\rangle \d v
\end{equation}
where angle brackets denote the spatial average.
Phasestrophy evolution is linked to the wave energy evolution, by a "$W$-$\Psi$ theorem",
\begin{equation}
\frac{\d W}{\d t} \,+\, 2 \gamma _d W  \;=\; \sum _s \frac{m_s u_s}{\d _v f_{0,s}} \left( \gamma ^\mathrm{col}_{\Psi} + \frac{\d}{\d t} \right) \Psi _s , \label{eq:phasestrophy_equation_pb}
\end{equation}
where $W=n_0 q^2 \left\langle E^2 \right\rangle /(m\omega _p^2)$ is the total wave energy, including sloshing energy. In the BB case, $u_s=\omega_p / (2k)$.
The above relation assumes that $f_{0,s}$ has a constant slope in the velocity-range spanned by evolving phase-space structures.
In parallel with quasi-geostrophic fluids, this relation is the kinetic counterpart of the Charney-Drazin non-acceleration theorem \cite{charney61}. 
Fig.~\ref{fig:bb_ps_growth} shows good quantitative agreement between the lhs and the rhs in a BB simulation. Fig.~\ref{fig:cdia_ps_growth} shows qualitative agreement between the lhs and the rhs in a CDIA simulation

In the BB case, we can apply the above general theory to obtain an expression for the nonlinear growth rate of an isolated phase-space structure,
\begin{equation}
\d \Psi /\d t = \left( \gamma _{\Psi}  - \gamma ^\mathrm{col}_{\Psi} \right) \Psi ,
\end{equation}
where $\gamma _{\Psi}$ is the collisionless phase-space structure growth-rate,
\begin{equation}
\gamma _{\Psi} \;\approx \; \frac{16}{3\sqrt{\pi}} \, \frac{\Delta v}{v_R} \, \frac{\gamma _{L0}}{\omega _p} \, \gamma _d. \label{eq:growthrate}
\end{equation}

Subcritical instabilities have been observed in BB simulations \cite{berkbreizman99,lesur09,lesur10bpsi} and current-driven ion-acoustic (CDIA) simulations \cite{berman82}.
Eq.~(\ref{eq:growthrate}) shows that the growth of structures is independent of linear stability, since it is not related to the sign of the total linear growth rate $\gamma \approx \gamma _{L0}-\gamma _d$. Nonlinear growth requires a positive $\gamma _d$ to enable momentum exchange, a positive slope for $f_0$ to provide free energy, and a seed structure with a width $\Delta v$ large enough for $\gamma _{\Psi}$ to overcome collisions. When the linear growth rate $\gamma$ is negative, the seed structure is the hole (clump) corresponding to the $v>v_R$ ($v<v_R$) part of the plateau, which is formed by particles trapped in the finite initial electric field.
We explain the mechanism of subcritical instabilities as follows. Landau damping generates a seed phase-space structure, whose growth rate can be positive if the growth due to momentum exchange overcomes decay due to collisions.
Subcritical instabilities have also been explained in terms of a nonlinear reduction of ion Landau damping by particle trapping \cite{nguyen10prl,nguyen10ppcf}, which is a different mechanism.

\begin{figure}
\includegraphics[scale=1]{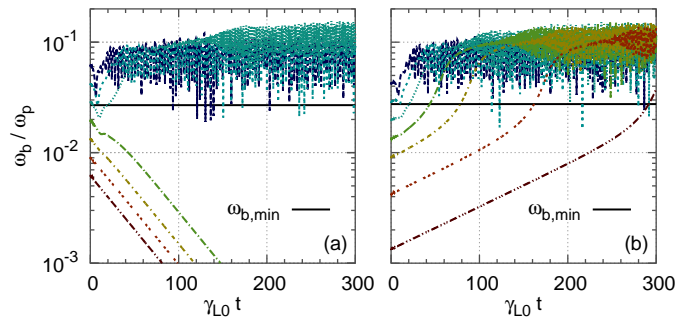}
\caption{\label{fig:threshold}Dashed curves: time-series of electric field amplitude for different initial amplitudes. (a) Subcritical case, $\gamma _d/\gamma _{L0}=1.05$. (b) Supercritical case, $\gamma _d/\gamma _{L0}=0.98$. Solid line: theoretical nonlinear instability threshold, Eq.~(\ref{eq:threshold}).}
\end{figure}

If Krook-like collisions are negligible, the initial amplitude threshold $\omega_{b,\mathrm{min}}$ is of the order of
\begin{equation}
\left( \frac{\omega_{b,\mathrm{min}}}{\omega _p}\right) ^2 \;\sim \; 0.12 \left( \frac{\omega _p}{\gamma _{L0}} \, \frac{\omega _p}{\gamma _d} \right) ^{2/3} \, \left( \frac{\nu _d}{\omega _p} \right) ^2. \label{eq:threshold}
\end{equation}
Fig.~\ref{fig:threshold}(a) shows time-series of electric field amplitude $\omega_b$ for different initial amplitudes, for the case $\gamma _d/\gamma _{L0}=1.05$, which is a subcritical instability with $\gamma/\gamma _{L0}=-0.045$. The threshold between damped solutions and nonlinear instabilities is in agreement with Eq.~(\ref{eq:threshold}).

In addition, our theory predicts the persistence of nonlinear instability in the marginally linear unstable regime.
The nonlinear instability due to phasestrophy growth is stronger than the linear growth if $ \gamma _{\Psi} - \gamma ^\mathrm{col}_{\Psi} > (3/2)\gamma$. We discovered numerically the existence of such supercritical nonlinear instabilities for $0<\gamma / \gamma _{L0} < 0.04$.
Fig.~\ref{fig:threshold}(b) shows time-series of electric field amplitude $\omega_b$ for different initial amplitudes, for $\gamma _d/\gamma _{L0}=0.98$, which is slightly above marginal stability with $\gamma/\gamma _{L0}=0.018$. The threshold where the linear growth becomes nonlinear is in agreement with Eq.~(\ref{eq:threshold}).

\section{Phase-space turbulence}

In the presence of multiple resonances, we observed avalanches in velocity-space due to the evolution of phase-space structures. Several holes and clumps emerge from neighboring resonances and interact with each other. Ultimately, holes coalesce into macro-scale structures, whose lifetimes are much larger than the classical quasilinear diffusion time and which thus dominate the nonlinear evolution. This finding reinforces the need for theoretical efforts toward a comprehensive theory of phase-space turbulence.

\begin{figure}
\includegraphics[scale=0.35]{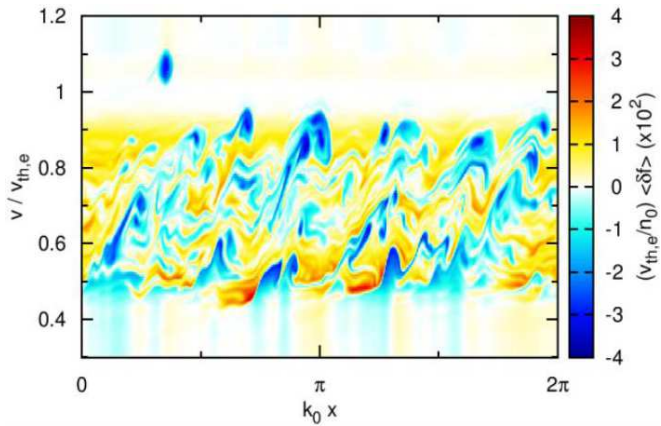}
\caption{\label{fig:PSturb}Phase-space turbulence. Snapshot of the electron phase-space in a CDIA simulation.}
\end{figure}

When the evolution of ions and electrons are accounted for, the CDIA can be excited in addition to Langmuir wave. The CDIA is linearly unstable when the velocity drift between ions and electrons exceeds some threshold. Even when the drift is much below this threshold, we found that wave energy is driven nonlinearly by phase-space structures, as seen in Fig.~\ref{fig:PSturb}, but not by high-level random noise. We found that phase-space structures have a significant effect on anomalous resistivity. When structures are negligibly small compared to the equilibrium particle distribution, but present in large number, they can have significant impacts, collectively. This was reported in Ref.~\cite{lesur12jspf}.

\section{Conclusions}

Our results are particularly relevant in the context of burning plasma. Indeed, a major concern is that high energy ions can excite plasma instabilities in the frequency range of Alfv{\'e}n Eigenmodes (AEs), which significantly enhance their transport. Transport and loss of fast particles depend on both the nonlinear saturation amplitude and the kind of nonlinear behaviour. Many qualitatively different nonlinear regimes have been observed in experiments \cite{fasoli97}, including a zoo of spectral components whose frequency shifts on a time-scale much smaller than profiles evolution time-scale (nonlinear chirping). The behaviours of these spectral components are qualitatively diverse in terms of their intermittency \cite{kusama99,podesta10,chen10}, their monotonicity in frequency shift \cite{berk06,nazikian08}, their asymmetry \cite{berk06,toi11}, and whether frequency shifting branches end as a continuous mode \cite{gryaznevich04} or not.
Near the resonant surface, it is possible to obtain a new set of variables in which the three-dimensional (3D) plasma is described by a one-dimensional Hamiltonian in two conjugated variables \cite{lichtenberg,berkbreizman97ppr,berkbreizman97pop,wong98,garbet08}, if we assume an isolated single resonance. In this sense, the problem of AEs is homothetic to the well-known paradigm of a single mode bump-on-tail instability. Observed quantitative similarities between BB nonlinear theory and both global simulations \cite{wong98,pinches04S47} and experiments \cite{fasoli98,heeter00,lesur_thesis,lesur10} are an indication of the validity of the aforementioned reduction of dimensionality.

In future work, we will investigate the existence and impacts of phase-space structures in ITG turbulence, in both theory/simulation, and in experiment with simplified geometry \cite{miwa13} . In particular the trapped-ion mode is a good candidate to detect phase-space structure dynamics.

\acknowledgments

I am grateful to my co-authors, and for stimulating discussions with D.~Escande, S.~Sharapov, B.N.~Breizman, M.K.~Lilley, and the participants in the 2009, 2011 and 2013 Festival de Th{\'e}orie. This work was supported by two grants-in-aid for scientific research of JSPF, Japan (21224014 and 25887041), by the collaboration program of the RIAM of Kyushu University and Asada Science Foundation, by the WCI Program of the NRF of Korea funded by the Ministry of Education, Science and Technology of Korea [WCI 2009-001], and by CMTFO via U.S.~DoE Grant No.~DE-FG02-04ER54738. Computations were performed on the Kraken system at NFRI, and on the XT system at Kyushu University.

%\bibliographystyle{prsty}
%\bibliography{tout}

\end{document}